\documentclass[tighten]{emulateapj}
\usepackage[]{times, graphicx}
\newcommand{\sdo}{{\em SDO{}}}
\newcommand{\hinode}{{\em Hinode}}
\newcommand{\alfven}{Alfv\'{e}n}
\newcommand{\alfvenic}{Alfv\'{e}nic}
\newcommand{\pref}{\protect\ref}
\newcommand{\sone}{$\mathrm{\sigma}_{{\mathrm{1/e}}}$}
\newcommand{\snt}{$\mathrm{\sigma}_{{\mathrm{nt}}}$}
\newcommand{\vrms}{$\mathrm{V}_{{\mathrm{RMS}}}$}
\newcommand{\vpp}{$\mathrm{V}_{{\mathrm{P2P}}}$}

\begin{document}

\shorttitle{Coronal ``Dark'' Energy}
\shortauthors{McIntosh \& De~Pontieu}
\title{Estimating the ``Dark'' Energy Content of the Solar Corona}

\author{Scott W. McIntosh\altaffilmark{1,3}, Bart De Pontieu\altaffilmark{2}} 
\altaffiltext{1}{High Altitude Observatory, National Center for Atmospheric Research, P.O. Box 3000, Boulder, CO 80307}
\altaffiltext{2}{Lockheed Martin Solar and Astrophysics Lab, 3251 Hanover St., Org. A021S, Bldg. 252, Palo Alto, CA  94304}
\altaffiltext{3}{Corresponding Author: \email{mscott@ucar.edu}}

\begin{abstract}
The discovery of ubiquitous low-frequency (3-5mHz) \alfvenic{} waves in the solar chromosphere (with Hinode/SOT), and corona (with CoMP and the {\em Solar Dynamics Observatory}, \sdo) has provided some insight into the non-thermal energy content of the outer solar atmosphere. However, many questions remain about the true magnitude of the energy flux carried by these waves. Here we explore the apparent discrepancy in the resolved coronal \alfvenic{} wave amplitude ($\sim$0.5km/s) measured by the Coronal Multi-channel Polarimeter (CoMP) compared to those of the \hinode{} and the {\em Solar Dynamics Observatory} (\sdo{}) near the limb ($\sim$20km/s). We use a blend of observational data and a simple forward model of \alfvenic{} wave propagation to resolve this discrepancy and determine the \alfvenic{} wave energy content of the corona. Our results indicate that enormous line-of-sight superposition within the coarse spatio-temporal sampling of CoMP hides the strong wave flux observed by \hinode{} and \sdo{} and leads to the large non-thermal line broadening observed. While this scenario has been assumed in the past, our observations with CoMP of a strong correlation between the non-thermal line broadening with the low amplitude, low frequency \alfvenic{} waves observed in the corona provide the first direct evidence of a wave-related non-thermal line broadening. By reconciling the diverse measurements of \alfvenic{} waves we establish large coronal non-thermal linewidths as direct signatures of the hidden, or ``dark'', energy content in the corona, and provide preliminary constraints on the energy content of the wave motions observed.
\end{abstract}


\section{Introduction}
\alfven{} waves have been invoked as a driving force behind solar coronal heating and wind acceleration since their inception \citep{Alfven1947}. While \alfvenic{} waves have been detected in situ by several spacecraft in and out of the ecliptic \citep{Belcher1971}, in the solar atmosphere their presence had only been inferred through the large non-thermal broadening of coronal emission lines  \cite[e.g.,][]{Hassler1990, Hassler1994, Banerjee1998, Chae1998, Peter1999a, Peter1999b, Peter2000a, Moran2001, Moran2003,  Peter2003, Akiyama2005, Singh2006a, Singh2006b}. Only recently were they directly observed and measured in the chromosphere \citep{DePontieu2007a} with the Solar Optical Telescope \citep[SOT;][]{SOT} on the {\em Hinode} spacecraft \citep{Hinode} and in the corona with the Coronal Multi-channel Polarimeter \citep[CoMP;][]{Tomczyk2007,Tomczyk2008,Tomczyk2009}. A subsequent investigation with the {\em Solar Dynamics Observatory} Atmospheric Imaging Assembly \citep[AIA;][]{2011SoPh..tmp..172L} confirmed the ubiquity of coronal wave motions \citep[][]{2011Natur.475..477M}. 

Unfortunately, there is a significant discrepancy in the energy estimates between these three direct observations. The \alfvenic{} wave energy flux in Doppler motions of the 10747\AA{} \ion{Fe}{13} emission line observed by CoMP is $\sim$0.01~Wm$^{-2}$ \citep{Tomczyk2007}, some four orders of magnitude less than the 100~Wm$^{-2}$ required to balance the radiative losses of the quiet corona \citep{WithbroeNoyes1977} and/or accelerate the fast solar wind \citep{HansteenLeer1995}. However, the observations of SOT indicate that an \alfvenic{} wave energy flux of order 4~kWm$^{-2}$ is present in the chromosphere \citep{DePontieu2007a} and that even a very low transmission ($\sim$3\%) of these waves into the corona \citep[due to reflection off the transition region,][]{Cranmer2005} will result in an energy flux of the order of $\sim$100~Wm$^{-2}$ into the corona or solar wind. In addition, the presence of {\it at least} that much wave energy is readily observed in the quiescent (quiet and coronal hole) corona with SDO/AIA \citep[][]{2011Natur.475..477M}. 

This leaves us in a quandary: where is the missing wave energy in the CoMP observations? Interestingly, the coronal wave energy flux inferred from the SOT observations (and measured by AIA) is consistent with the amplitude of non-thermal broadening observed by CoMP, as reported by \citet{Tomczyk2007}. So, perhaps the real energy flux of \alfvenic{} waves in the corona is hidden from CoMP. In this scenario, the vastly different spatial scales of the instruments involved would result in considerable superposition of low frequency motions resolved by SOT and AIA but unresolved by CoMP, quenching the resolved velocity amplitude, energy flux, and increasing the non-thermal line broadening of the observed line profiles in the optically thin corona. An alternative scenario faces the challenging task of simultaneously explaining the large non-thermal broadening in the corona, and the observational evidence from AIA of strong \alfvenic{} waves throughout the corona. Perhaps the filling factor of the waves observed with AIA is much lower than estimated and/or the bulk of the \alfvenic{} waves (coming up from the chromosphere) are actually damped and/or dissipated in the first few thousand kilometers above the limb, leaving only low resolved wave amplitudes for CoMP? Perhaps the non-thermal broadening of the coronal emission line is then caused by unresolved flows into and out of the plane-of-the-sky (POS)? 

\begin{figure*}
%\epsscale{1.15}
%\plotone{f1}
\center
\includegraphics[width=170mm]{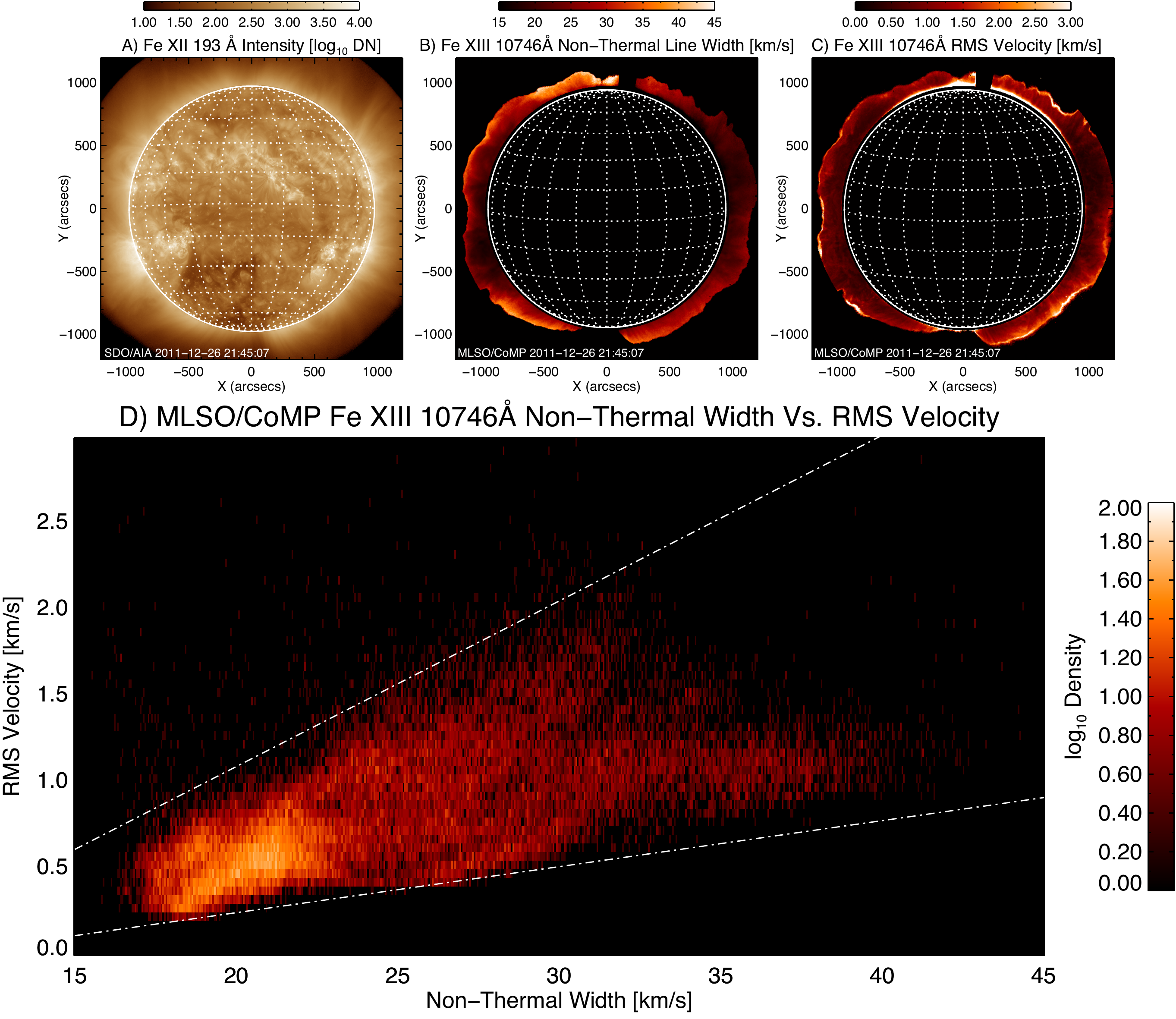}
\caption{Comparing the non-thermal linewidth (\snt) and root-mean-square of the Doppler velocity (\vrms) of CoMP \ion{Fe}{13} 10774\AA{} line with the AIA 193\AA{} from December 26, 2011. Supporting movies of the CoMP peak intensity, Doppler velocity, and non-thermal width can be found in the electronic edition of the Journal. \label{f1}}
\center
\end{figure*}

In this Letter we investigate these possible scenarios, address the apparent mis-match of \alfvenic{} wave energies observed by SOT, AIA, and CoMP and thus constrain the potential sources of non-thermal broadening, the invariant hidden, or ``dark'' energy, in the quiet corona. We expand upon a simple forward model \citep{McIntosh2008} that explained the center-to-limb variation of transition region lines by taking into account that SOT and AIA observed a lower atmosphere dominated by a myriad of extremely fine-scale, field-aligned and independently moving structures, called spicules, with spatial widths $<$300 km (considerably smaller than the $\sim$4,500 km CoMP spatial sampling) that exhibit significant transverse motions 
% BDP: what is the actual spatial sampling of CoMP? And spatial resolution? Is it Nyquist dominated? Or diffraction limited? There are various numbers in the paper
\citep{DePontieu2007a, 2011Natur.475..477M}. 
%The measured spicule properties \citep{DePontieu2007b} were employed to demonstrate that the distribution of non-thermal linewidths across the solar disk was consistent with the mixture of sub-resolution longitudinal and transverse motions where the latter needed to have significant amplitudes ($\sim$20-30 km/s) to reproduce the observed behavior.
% - this forward model forms the basis of the following work. 

%A forward model of a corona riddled with \alfvenic{} motions must simultaneously recover the resolved velocity amplitude and the apparent temporal invariance of the non-thermal line broadening in order to offer insight. 
In the following section we discuss the observations that drive the model. In Sect.~\pref{method}, we discuss how the SOT and AIA observations are used to drive the model. In Sect.~\pref{analysis} we look at the properties of the model that can reproduce the observed CoMP behavior.

\section{Observations}\label{obs}
\subsection{CoMP Observations}\label{Cdata}
%
%The Coronal Multi-channel Polarimeter \citep[CoMP;][]{Tomczyk2008} is a combination polarimeter and narrowband tunable filter that can measure the complete polarization state in the vicinity of the 10747\AA{} and 10798\AA{} \ion{Fe}{13} coronal emission lines. It was deployed behind the 20-cm aperture Coronal One Shot (COS) coronagraph \citep{Smartt1981} and is now mounted on the spar at the Mauna Loa Solar Observatory (MLSO). CoMP is comprised of: 1) an occulting disk, located at the focus of the COS, that blocks the light from the solar disk; 2) a lens that collimates the solar image; 3) a filter wheel holding three order-blocking filters corresponding to each of the three observable emission line regions; 4) the polarimeter/tunable filter package; 5) a re-imaging lens that forms the final solar image; and 6) a 1024x1024 pixel HgCdTe infrared detector array. The CoMP filter is a four-stage, wide-field calcite birefringent filter with a bandwidth of 1.3\AA{} (an instrumental width of 21km/s) and is tuned in wavelength by four liquid crystal variable retarders and has a full field of view of 2.8R$_{\sun}$ at a spatial sampling of 4.5\arcsec.

The CoMP observations used in this paper were obtained on December 26, 2011 from 20:16 to 21:45UT in three wavelengths (10745.2, 10746.5, and 10847.8\AA{}) across the 10747\AA{} \ion{Fe}{13} line, with an exposure time of 250ms at each position. The image groups were obtained at a cadence of 30 seconds. 
%Each image is reduced with the following recipe: 1) subtract a dark image; 2) divide by an image taken with a calibrated diffuser in front of the objective lens in order to remove pixel-to-pixel detector gain variations, normalize the relative transmission of the filter bandpasses, and normalize the intensity into units of the solar disk central intensity; 3) determine the location of the images on the detector, translate to a common center, and rotate to orient solar north up; 4) subtract continuum from line. 
From the three wavelength positions we measure the peak line intensity, Doppler shift, and linewidth of the 10747\AA{} line. Fitting a histogram to the measured velocities allows us to determine the coronal rotation rate and estimate the ``rest'' wavelength to be 10746.08\AA{} (assuming that half of the field of view is red and half is blue\--shifted). For an instrumental width ($\sigma_{inst}$) of 21km/s and a thermal width ($\sigma_{th}$) of 21km/s for \ion{Fe}{13} (using $\sigma_{th} = \sqrt{2k_{B}T_{e}^{\ast} / m_{ion}}$, for an ion of mass $m_{ion}$ and peak formation temperature of $T_{e}^{\ast}$ assuming that ion and electron temperatures are equal) we determine the non-thermal width \snt{} $(=  \sqrt{\sigma_{1/e}^2 - \sigma_{inst}^2 - \sigma_{th}^2})$, where $\sigma_{1/e}$ is the 1/e width of the fitted line profile.

The top row of Fig.~\pref{f1} compares an AIA 193\AA{} image (panel A) taken at the start of the timeseries with the CoMP \snt{} (panel B) and root-mean-square of the Doppler velocity during the timeseries (\vrms; panel C). The latter is dominated by the resolved, low-frequency, and low-amplitude \alfvenic{} waves that are seen to propagate in the corona. Panel D shows a two-dimensional histogram of \vrms{} and \snt{}. We find a strong correlation between \vrms{} and \snt{}. This is an indication that the two measures are intimately tied, i.e., the variation of the resolved wave amplitude during the timeseries and the non-thermal broadening of the line in a pixel/voxel are co-dependent. This suggests that whatever causes the strong non-thermal line broadening is related to the \alfvenic{} waves that CoMP observes. Note the clear cut-off at \snt{}=17-18km/s and the enclosing ``correlation wedge'' of the two dot-dashed lines. This wedge is a robust feature of the CoMP observations, shifting slightly left and right and tilting slightly up and down from day to day as the physical structure of the corona above the limb changes. Supporting movies of SDO/AIA (\ion{Fe}{12}) 193\AA{}, CoMP I, V, and $\sigma_{1/e}$ can be found in the electronic edition of the Journal, the latter clearly illustrate the ubiquity of the transverse Doppler motions. %The relationship shown in panel D motivates the remainder of the present investigation, as it suggests that the non-thermal broadening holds information about the true \alfvenic{} energy content of the corona. 

\subsection{Hinode/SOT Observations}\label{Hdata}
%
%SOT observes the photosphere and chromosphere with unprecedented spatial ($\sim$0.2\arcsec) and temporal ($\sim$5 s) resolution. 
Hinode/SOT observations have shown that the chromospheric limb is dominated by superposition of a large number of long, thin ($\le$200km) spicules that extend to heights of 5-15\arcsec{} and that undergo significant motions along and perpendicular to the line-of-sight (LOS). \citet{DePontieu2007b} demonstrated that there are two types of spicules present in the {\em Hinode} limb observations.
% with very different dynamic properties. 
Type I spicules evolve on timescales of 3-5 minutes reaching velocities of 10-40km/s along their long axis \citep{DePontieu2007a}. Type II spicules on the other hand, occur on timescales of 10-120s and reach apparent longitudinal velocities of 50-150km/s. \cite{DePontieu2007b} have shown, using a statistical approach, that these chromospheric features undergo vigorous \alfvenic{} motions with amplitudes around $20(\pm5)$km/s and periods ranging from 100 to 500s.

\subsection{SDO/AIA Observations}\label{Adata}
\citet{2011Natur.475..477M}, using a similar approach, demonstrated that an energetically relevant portion of the \alfvenic{} energy observed in the chromosphere by SOT is indeed visible in emission formed at coronal temperatures. Those waves have made it ``into'' the corona \-- these observations were made possible by the high S/N, low scattered light, high spatial resolution observations of AIA. The imaged wave motions have amplitudes, periods, and speeds commensurate with those observed by SOT near the limb, and can be considered the coronal extension \citep{DePontieu2011} of the chromospheric features studied by \cite{DePontieu2007b}. The observed phase speeds of these coronal \alfvenic{} waves reach 500\--1000km/s, values that are commensurate with those observed by CoMP \citep[][]{Tomczyk2007,Tomczyk2009}. 

\begin{figure*}
%\epsscale{0.9}
%\plotone{f2}
\center
\includegraphics[width=170mm]{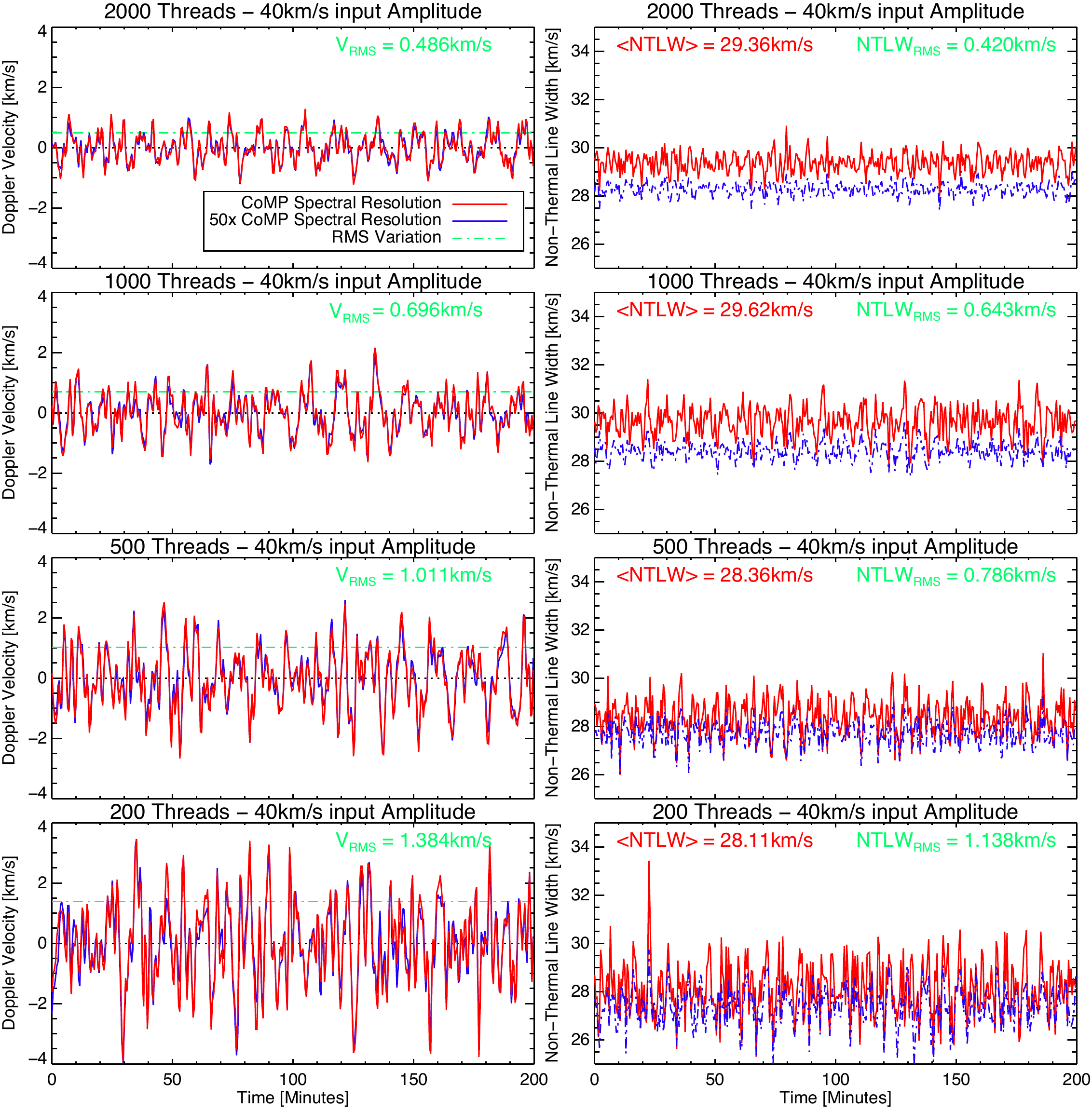}
\caption{Sample timeseries for different numbers of emitting threads in a coronal voxel (increasing from bottom to top: 200, 500, 1000, and 2000) for the three-point CoMP measurements (red) and an imaginary spectrograph with 50 times the spectral sampling (blue). The left column of panels show the variation of the Doppler velocity and \vrms{} (green dot\--dashed line) while the right column shows the variation of \snt{} for a sample timeseries of 200 minute duration. \label{f2}}
\center
\end{figure*}

\section{Method \& Analysis}\label{method}
We use the high resolution SOT \citep[][]{DePontieu2007a} and AIA \citep[][]{2011Natur.475..477M} observations as a guide for the transverse motions that occur on a spatial scale that is far smaller than CoMP resolution. We develop a (simple) Monte-Carlo forward model of \alfvenic{} waves in the corona based on the concept of a ``thread'', or elementary oscillating structure. Our simple picture is that any pixel in the optically thin model corona can be populated with $N$ independent threads and the 10747\AA{} signal observed is given (at any time) by summing the emission of all of the threads in any given pixel. 

The properties of any single thread are as follows: a thread can appear at any time with a lifetime chosen from a Gaussian distribution of 100 $\pm$ 20s; has a uniform brightness; and is subject to \alfvenic{} motion governed by a randomly chosen period, amplitude and phase. Here we consider swaying motions, since those are clearly observed with AIA. We assume a polarization angle chosen from a uniform distribution between 0 and $2\pi$ with respect to the LOS, and apply a $1 / \cos \theta$ correction factor to accommodate the fact that not all threads are radial (where $\theta$ follows a Gaussian distribution of 0$\pm$20\degr). The sinusoidal \alfvenic{} motions of the thread have a period drawn from a distribution based on the CoMP power spectrum \citep[see, e.g., Fig.~2 of][]{Tomczyk2007}, a phase from a uniform distribution between 0 and $2\pi$ and an amplitude from a Gaussian distribution around X$\pm$5 km/s (where X can be any value $>$5km/s). Each thread emits a Gaussian line profile of width 21km/s (equivalent to $\sigma_{th}$ of the \ion{Fe}{13} plasma) at the given LOS velocity determined from its period, amplitude and phase. The resulting line profile in a pixel is the sum of the Gaussians of the $N$ individual threads.%, and reflects the ensemble motion of the coronal plasma threads within the pixel.

\begin{figure}
%\epsscale{0.9}
%\plotone{f3}
\center
\includegraphics[width=85mm]{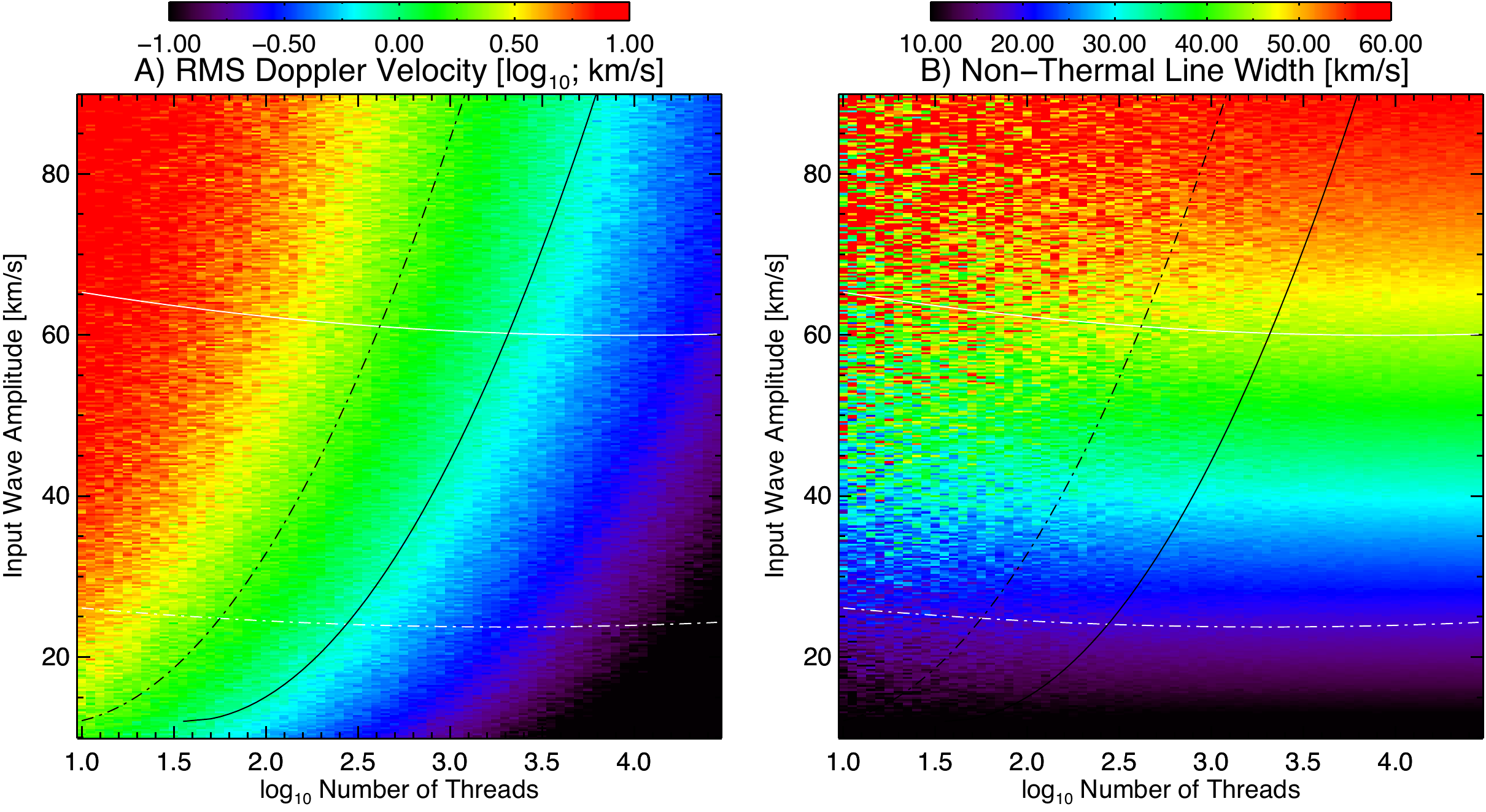}
\caption{Variation of \vrms{} (A) and \sone{} (B) measured from the Monte-Carlo simulated timeseries of three-point 10747\AA{} measurements for a range of thread numbers and input wave amplitudes. The white (dot-dashed and solid) lines show the 15 and 45km/s values of \snt{} respectively, while the black (dot-dashed and solid) lines correspond to the 0.5 and 1.5km/s values of \vrms{} respectively. \label{f3}}
\center
\end{figure}

We then observe these synthetic profiles in the same fashion as the CoMP observations \citep{Tomczyk2007} by sampling the 10747\AA{} line at the three wavelengths above with a Gaussian filter width of 1.3\AA{}, and fitting the ensemble line profiles with a single Gaussian to derive I, V, and \sone{}. Figure~\pref{f2} shows sample (pixel) timeseries of the Doppler velocity (and \vrms; left column) and non-thermal linewidth (\snt{}; right column) for a range of thread numbers (bottom to top: 200, 500, 1000, and 2000) and a fixed 40km/s wave (peak-to-peak) amplitude (\vpp). Notice the diminishing values of \vrms{} with increasing thread number \-- these values eventually saturate at the Gaussian fitting accuracy of the three-point measurements ($\sim$0.1 of the spectral resolution). Notice also that \snt{} $\approx \sqrt 2$\vpp{} and the gradual increase in \snt{} with increasing thread number and the reduction of its variance with time \-- recalling that \citet{Tomczyk2007} noticed an invariance of the linewidths with time in their original measurements. To investigate any artifacts introduced by the three-point measurements (red in Fig.~\pref{f2}), we also perform the same calculations of velocity and linewidth using synthetic data at considerably higher spectral resolution (blue). We find that there is no significant difference in \vrms{} or \snt{} for both spectral resolutions except in the case of large thread numbers where the profiles can become non-Gaussian. 
%This gives some comfort that the three-point measurements of CoMP can adequately describe the centroid shift (second moment) of the 10747\AA{} line. 
%The spectral resolution does affect the determination of the non-thermal linewidths, with the three-point measurement (red) consistently underestimating the non-thermal line broadening (as observed at super-resolution, blue). %This is driven by the improved sampling of the line profile.
In summary, it is clear from Fig.~\pref{f2} that our model can reproduce both the low amplitudes of the resolved \alfvenic{} waves and large {\em temporally invariant} non-thermal line broadening observed with CoMP.

Extending this computation to a broader range of models is straightforward. The panels of Fig.~\pref{f3} show the variation in \vrms{} (A) and \sone{} (B) for a range of thread numbers and input wave amplitudes where the other thread parameters remain as described above. The white (dot-dashed and solid) lines show the 15 and 45km/s values of \snt{} respectively, while the black (dot-dashed and solid) lines correspond to the 0.25 and 1.5km/s values of \vrms{} respectively. These bounds encapsulate more than 75\% of the analyzed pixels in panels B and C of Fig.~\pref{f1}. Note that iso-contours of \vrms{} are approximately quadratic functions of the (log of the) thread number and input wave amplitude, whereas 
%while the largest sensitivity of \sone{} (and hence \snt) to the variables is for input wave amplitudes of order 35km/s and thread numbers below 1000. 
%BDP: I don't really understand the above, hence deleted it -- not sure it's relevant, and may just confuse
the measured linewidths almost directly reflect the input wave amplitude of the model (for thread numbers larger than $\sim$30).

\begin{figure}
%\epsscale{0.75}
%\plotone{f4}
\center
\includegraphics[width=85mm]{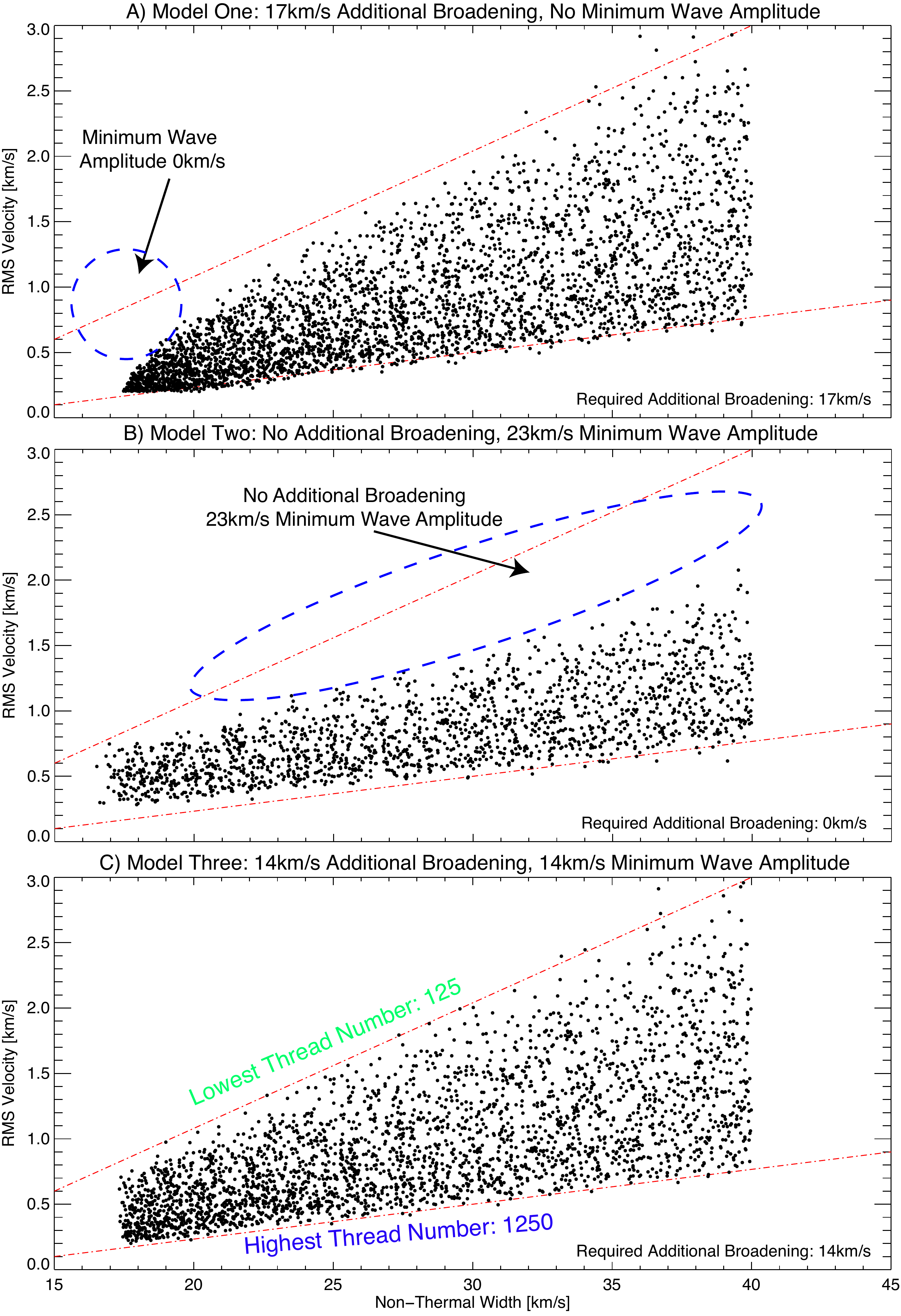}
\caption{Scatter plots of the \vrms{} and \snt{} values used to reproduce the variance observed in panel D of Fig.~\pref{f1}. The red dashed lines are taken from  the bounds of the correlation wedge in Fig.~\pref{f1}. Panel A has no minimum wave amplitude, minimum and maximum thread numbers of 110 and 1200 (respectively), but requires 17km/s of additional broadening to match the observed cut-off velocity of the wedge. Panel B has non additional broadening, minimum and maximum thread numbers of 300 and 1500 (respectively), but requires 23km/s minimum wave amplitude to match the observed cut-off velocity of the wedge. Panel C is the optimal solution with 14km/s additional broadening, a 14km/s minimum wave amplitude and thread limits as shown. \label{f4}}
\center
\end{figure}

One of the most stringent constraints from CoMP for our model is the correlation and peculiarly shaped correlation wedge that we found in Fig.~\pref{f1}D between the \vrms{} and \snt{}. We show in Fig.~\pref{f4} that we can easily (and naturally) reproduce such a correlation of \vrms{} and \snt{} and the wedge by using the data from the simple model presented in Fig.~\pref{f3}. We use the line-enclosed regions in the panels of Fig~\pref{f3} (which are driven by CoMP measurements), and apply other simple selection rules on the parameter set to best recover the correlation wedge. We find that the shape of the wedge is sensitive to the number of threads along the line of sight, the minimum wave amplitude and any additional broadening not due to swaying motions. Panel A shows the relationship between \vrms{} and \snt{} when the minimum wave amplitude is allowed to be zero. To match the cut-off of the wedge at 17km/s shown in the observations of Fig.~\pref{f1} we need to add 17km/s of additional broadening. The maximum and minimum number of threads determines the lower and upper slopes of the wedge. While the slopes of the wedge are generally well reproduced, the low end of the wedge is too pointed and the scatter in \vrms{} does not fill the wedge. This implies that this parameter set is inadequate to reproduce the observed relationship. Similarly, in panel B, we have omitted the additional broadening but, to match the observed low-end cut-off at 17km/s we must impose a minimum wave amplitude of 23km/s peak-to-peak (which is equivalent to 16km/s rms). This parameter set is also inadequate: the slope of the wedge is too shallow and it does not fill the wedge at values of high \vrms{} and \snt{}. Finally, in panel C, we show the optimal parameter set for this observation. The best match fills the wedge and requires a minimum wave amplitude of 14km/s peak-to-peak (which is equivalent to 10km/s rms) and an additional broadening of 14km/s rms. The high \snt{} cut-off would indicate that the maximum wave amplitude present is 40km/s (or 56km/s peak-to-peak). The upper bound of the wedge is determined by pixels that have low thread numbers where we find the best match is at 125 threads per voxel. The lower bound of the wedge is determined by the highest number of threads in a voxel where the best match is between 1200 and 1300 threads. In summary, the correlation found from CoMP comes about naturally if the observations are dominated by superposition of vigorous \alfvenic{} motions (with properties similar to SOT and AIA) along a LOS with a large and variable number of independently oscillating structures.

\section{Discussion}\label{analysis}
To understand the low values of the resolved CoMP wave amplitude we have constructed a simple Monte-Carlo model with only two variables: the input wave amplitude, and the number of emitting threads in a voxel. Our model can reproduce the CoMP observations, even though it is based on the much larger wave amplitudes observed with SOT and AIA: CoMP measurements will see decreasing wave amplitudes and increasing linewidths with increasing complexity (superposition) of the emission along the LOS. We also see that the linewidths approach temporal invariance within the accuracy of the Gaussian fit - this invariance was noted by \citep{Tomczyk2007}. We note that the movies supporting Fig.~\pref{f1} show that in the less dense regions (in coronal holes), the linewidths do show some periodicity with time. This would appear to be consistent with the lower panels of Fig.~\pref{f2} which show low amplitude variations of linewidths where the number of threads is smaller. The earlier deduction of complete temporal invariance \citep{Tomczyk2007} may have been biased by the study of a single sub-region.

\begin{figure*}
%\epsscale{0.9}
%\plotone{f5}
\center
\includegraphics[width=170mm]{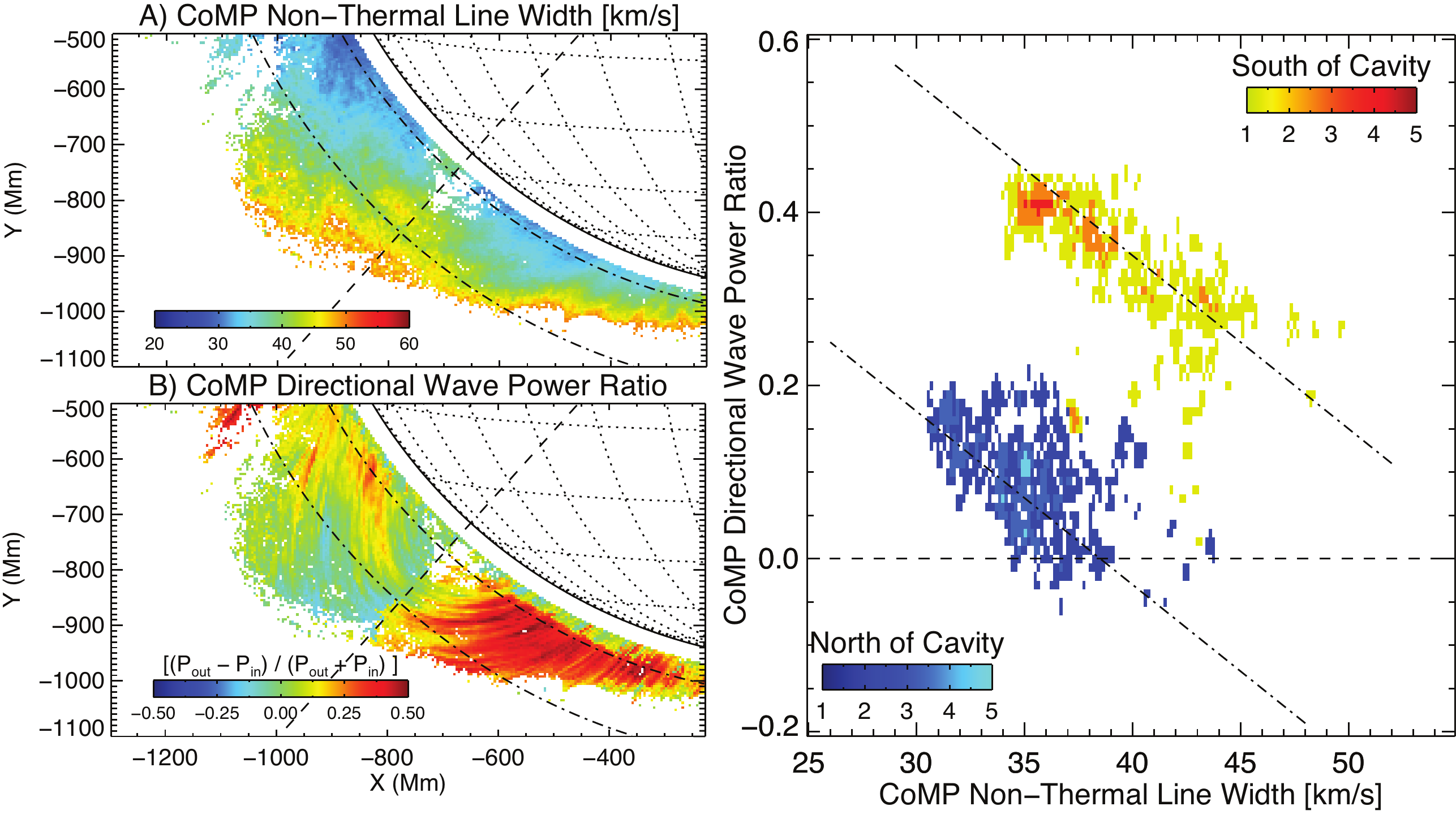}
\caption{Comparing \snt{} (panel A) and directional wave power ratio (panel B) from CoMP in the coronal cavity where the wave signal to noise is the highest as studied by \citet{Tomczyk2007} and \citet{Tomczyk2009}. The cavity is divided into a North and South component by the dashed radial line. The right panel shows the correspondence between these quantities for both portions of the coronal cavity. \label{f5}}
\center
\end{figure*}

We have also reproduced the wedge relationship between \vrms{} and \snt{}. This relationship comes about naturally if one limits the input wave amplitudes and assumes a minimum and maximum number of independently oscillating threads along the LOS. Under the assumptions of our simplistic model, this would imply that observed CoMP wave amplitudes of $\sim$1km/s are actually hiding the true wave amplitudes that range between 25 and 56km/s. Further, the measured linewidths show an almost one-to-one relationship with the amplitude of the waves present (Fig~\pref{f3}B), thus providing, in principle, an almost direct measure of the unresolved low-frequency wave energy. We note that the exact calibration between line broadening and wave amplitude of swaying motions is actually dependent on several assumptions (see below).

Unresolved flows along the LOS likely contribute to the non-thermal line broadening observed with CoMP. In principle, such flows could even cause the correlation found in Fig.~\pref{f1}D, as long as the flows are temporally varying (\vrms). However, this is unlikely for a couple of reasons. First, the \vrms{} is not dominated by flows but by waves \citep[][]{Tomczyk2007}. Second, as shown in Fig.~\pref{f5}, the upward-downward directed wave power ratio of \citep[][]{Tomczyk2009} shows the same correlation with the \snt{} as \vrms{}. Dividing the coronal cavity into its southern (foreground) and northern (background) portions by the dashed radial line we plot the correlation of \snt{} and the directional wave power ratio. In both regions the relationship is linear, which supports the scenario in which \snt{} is dominated by the propagating wave motions of the plasma and not by bulk coronal flows along the LOS. 

Our simplistic model has several limitations. For example, we assume that the transverse motions are dominated by swaying, and not torsional, motions. The latter would not cause resolved Doppler shift variations with time. It is possible that the additional broadening of 14km/s (uncorrelated with \vrms) required to reproduce the correlation wedge is a signature of such torsional motions. This would be consistent with recent observations of peak-to-peak amplitudes in spicules of order 20-30 km/s \citep[][]{2012ApJ...752L..12D}. It is also possible, as recent studies suggest, that some of the swaying motions couple to torsional modes \citep{{Verth2010,2011ApJ...731...73P,2012ApJ...746...31D}}. If such a coupling mechanism occurs in the corona \citep[see, e.g.,][]{Ionson1978,Ruderman2002}, the amplitude of the torsional modes could be correlated with that of the swaying motions, so that the line broadening caused by torsional modes could actually be part of the broadening that {\em is} correlated with \vrms. In this case, torsional motions would {\em not} cause the additional broadening (which instead could be caused by unresolved flows along the LOS). If this were the case, the observed correlation between \vrms{} (from swaying) and \sone would not necessarily all be caused by LOS superposition.
%This is unlikely to be the case at lower heights ($<$50Mm), where AIA shows evidence of swaying motions of large amplitude, but perhaps torsional motions play an increasingly larger role at greater heights. 
%Detailed, time-dependent measurements at high resolution will be required to further constrain the relative role of swaying and torsional wave motions.
In this scenario, the relative contributions of swaying and torsional motions likely change with height (with increasing contributions from torsional motions higher up), because the coupling mechanism suggested by \citet{2012ApJ...746...31D} occurs over a finite time /distance during which waves propagate upward from the loop footpoints.

Taken at face value, our model would predict a very large number of independently oscillating threads along the LOS ($\sim$1 solar radius) contained in one CoMP pixel (4,500 km). It may well be the case that the corona is extremely finely structured. However, the number of threads predicted by our model may be very different from the number of threads in the corona. 
%for example we have assumed uniform material density on the threads, but it may be that the processes governing the LOS profile production are linear enough that simply adding more threads can compensate for the inaccuracy of our initial density estimate. 
%Or, are we saying that the corona is extremely finely structured \-- if {\em only} low frequency waves are responsible for the line broadening then we need a lot of superimposed structure to do that \-- 
This is because the exact number will critically depend on the mix of torsional and swaying motions, and on the (perpendicular) coherence of the waves. The latter is unknown: if phase mixing plays a significant role, wave amplitudes seen by CoMP can, in principle, be quenched much more efficiently than the completely independently oscillating threads assumed here. 

Our data seem to exclude a scenario whereby high frequency \alfvenic{} waves (invoked by several models for the solar wind) dominate non-thermal line broadening (at least for heights $<$50Mm). If the wave generation process low down predominantly created high frequency waves, with a small amount of wave power at low frequencies, one could in principle recover the correlation wedge. However, the AIA and SOT observations of strong, low frequency \alfvenic{} swaying at low heights argue against such a scenario.

\acknowledgements
NCAR is sponsored by the National Science Foundation. CoMP data can be found at \url{http://mlso.hao.ucar.edu/}. {\em Hinode} is a Japanese mission developed and launched by ISAS/JAXA, with NAOJ as a domestic partner and NASA and STFC (UK) as international partners. We acknowledge support from NASA contracts NNX08BA99G, NNX11AN98G, NNM12AB40P, NNG09FA40C ({\em IRIS}), and NNM07AA01C (\hinode).

%\clearpage

\end{document}